\documentclass[10pt,aps,prd,nofootinbib,superscriptaddress,twocolumn,preprintnumbers,balancelastpage]{revtex4}
\pdfoutput=1
\usepackage{latexsym}
\usepackage{amssymb}
\usepackage{epsfig,amsmath,graphics}
\usepackage{listings}
\usepackage{color}
\usepackage{dcolumn}
\usepackage{slashed}
\usepackage{cancel}
\usepackage{comment,latexsym} 
\usepackage{bm}
\usepackage{verbatim}
\usepackage{tabularx}
\usepackage{dcolumn}
\definecolor{Mahogany}{rgb}{0.62,0.24,0.15}
\definecolor{colorLink}{rgb}{0.7,0,0}
\definecolor{colorCite}{rgb}{0,.7,0}
\definecolor{colorURL}{rgb}{0,0,0.7}
\definecolor{colorTC}{rgb}{.2,.7,.2}
\definecolor{colorDP}{rgb}{.7,.7,.2}
\usepackage[colorlinks=true,linktocpage=true,linkcolor=black,citecolor=colorCite
,urlcolor=colorURL]{hyperref}
\usepackage{setspace}
\usepackage{xspace}
\usepackage{multirow}
\usepackage{titlesec}
\usepackage[bbgreekl]{mathbbol}
\usepackage{bm}
\usepackage{float}
\usepackage{placeins}
\usepackage{afterpage}
\usepackage{booktabs}
\usepackage{pifont}

\graphicspath{{Figures/}}

\def\be{\begin{equation}}
\def\ee{\end{equation}}
\newcommand{\beq}{\begin{equation}}
\newcommand{\eeq}{\end{equation}}
\newcommand{\eref}[1]{Eq.~(\ref{#1})}

\newcommand{\lsim}{\!\mathrel{\hbox{\rlap{\lower.55ex \hbox{$\sim$}} \kern-.34em 
\raise.4ex \hbox{$<$}}}}
\newcommand{\gsim}{\!\mathrel{\hbox{\rlap{\lower.55ex \hbox{$\sim$}} \kern-.34em 
\raise.4ex \hbox{$>$}}}}

\definecolor{colorTC}{rgb}{.2,.7,.2}

\expandafter\def\expandafter\normalsize\expandafter{%
    \normalsize
    \setlength\abovedisplayskip{8pt}
    \setlength\belowdisplayskip{8pt}
    \setlength\abovedisplayshortskip{8pt}
    \setlength\belowdisplayshortskip{8pt}
}

\begin{document}

\title{What is the Machine Learning?\vspace{-2pt}} 

\author{Spencer Chang, Timothy Cohen, and Bryan Ostdiek\\[1pt] 
\emph{\small Institute of Theoretical Science, University of Oregon, Eugene, Oregon 97403}
\vspace{-4pt}
}

\begin{abstract}
\vskip -6 pt
\begin{center}
{\bf Abstract}
\end{center}
\vskip -7 pt
Applications of machine learning tools to problems of physical interest are often criticized for producing sensitivity at the expense of transparency.  To address this concern, we explore a \emph{data planing} procedure for identifying combinations of variables -- aided by physical intuition -- that can discriminate signal from background.  Weights are introduced to smooth away the features in a given variable(s).  New networks are then trained on this modified data.  Observed decreases in sensitivity diagnose the variable's discriminating power.   Planing also allows the investigation of the linear versus nonlinear nature of the boundaries between signal and background.  We demonstrate the efficacy of this approach using a toy example, followed by an application to an idealized heavy resonance scenario at the Large Hadron Collider.  By unpacking the information being utilized by these algorithms, this method puts in context what it means for a machine to learn.
$\quad$
\begin{spacing}{1.05}\noindent

\end{spacing}
\end{abstract}

\maketitle

A common argument against using machine learning for physical applications is that they function as a black box: send in some data and out comes a number.  While this kind of nonparametric estimation can be extremely useful, a physicist often wants to understand what aspect of the input data yields the discriminating power, in order to learn/confirm the underlying physics or to account for their systematics.  A physical example studied below is the Lorentz invariant combination of final state four-vectors, which exhibit a Breit-Wigner peak in the presence of a new heavy resonance.  

\begin{figure}[t!]
\vspace{-10pt}
\begin{center}
\includegraphics[width=  0.9\linewidth]{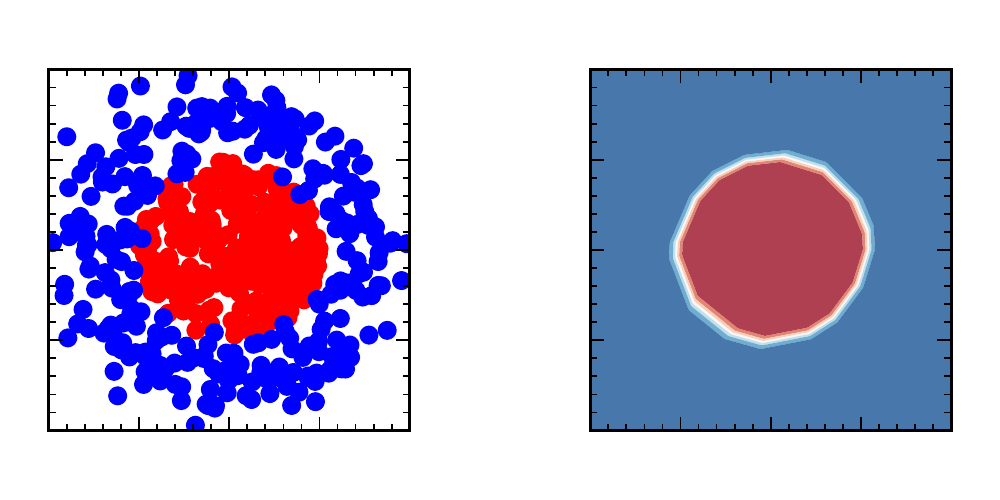}
\vspace{-10pt}
\caption{[Left] The machine is trained using rectilinear coordinates to distinguish blue and red as defined by the displayed training data. [Right]  The classifier output ranges from blue to red.}
\label{fig:Circle}
\vspace{-20pt}
\end{center}
\end{figure}

The simple example illustrated in Fig.~\ref{fig:Circle} exposes the subtlety inherent in extracting what the machine has ``learned."  The left panel shows red and blue data, designed to be separated by a circular border.  The right panel shows  the boundary between signal and background regions that the machine (a neural network with one hidden layer composed of 10 nodes) has inferred.   Under certain assumptions, a deep neural network can approximate any function of the inputs, \emph{e.g.}~\cite{Cybenko1989}, and thus produces a fit to the training data.  While any good classifier would find a ``circular" boundary, simply due to the distribution of the training data, one (without additional architecture) has no mechanism of discovering it is a circle. In light of this, our goal is to unpack the numerical discriminator into a set of human-friendly variables that best characterize the data.  While we are not inverting the network to find its functional form, we are providing a scheme for understanding classifiers.

For context, we acknowledge related studies within the growing machine learning for particle physics literature. The authors of~\cite{Baldi:2014kfa,Baldi:2014pta,Baldi:2016fql,Guest:2016iqz} emphasized the ability of deep learning to outperform physics inspired high-level variables.  We use the ``uniform phase space'' scheme to flatten in discriminating variables that was introduced in~\cite{deOliveira:2015xxd} to quantify the information learned by deep neural networks.  For other suggestions on testing what the machines are learning, see~\cite{Stevens:2013dya,Rogozhnikov:2014zea,Almeida:2015jua,  Kasieczka:2017nvn, Datta:2017rhs,  Aguilar-Saavedra:2017rzt}.  A nice summary of these ideas can be found in~\cite{Larkoski:2017jix}. Additionally, progress has recently been made in the related question of \emph{how} the machine learns \cite{2014arXiv1410.3831M, 2017arXiv170300810S}.

Section~\ref{sec:Proceedure} introduces a simple weighting scheme, which we call \emph{data planing}.\footnote{Planing is a woodworking technique for smoothing a surface.}  Applications to a toy model will be presented to illustrate the features of this approach.   As we demonstrate, it is possible to plane away all the underlying discriminating characteristics of this toy by utilizing combinations of linear and nonlinear variables.  This highlights another salient attribute of data planing: by comparing the performance of linear and deep neural networks, one can infer to what extent the encoded information is a linear versus nonlinear function of the inputs.  Then in Sec.~\ref{sec:PhysicsExample} we show that these features can be realized in a more realistic particle physics setting.  Finally, Sec.~\ref{sec:Discussion} concludes this paper with a discussion of future investigations.

\setcounter{section}{1}
\section{DATA PLANING}
\label{sec:Proceedure}
\vspace{-10pt}
Our starting assumption is that a sufficiently deep network with ample training can take advantage of all inherent information to discriminate signal from background, \emph{i.e.},  the network approximately attains Bayes error~\cite{Fukunaga19901, bayes2},  the lowest possible error rate.   The  approach advocated here is to then remove information, where the  performance degradation of the new networks provides diagnostic value (this procedure was first introduced in the ``uniform phase space'' section of \cite{deOliveira:2015xxd}).   To plane the data, we weight the events, which are labeled by $i$ and characterized by input variables $\vec{x}_i$.  After choosing a variable $m$, the planing weights are computed using 
\begin{align}
\Big[w\big(\vec{x}_i\big)\Big]^{-1}  = C\, \frac{\text{d} \sigma\big(\vec{x}_i\big)}{\text{d} m}\Bigg|_{m=m_i}\,,
\label{eqn:weight}
\end{align}
where $\text{d}\sigma/\text{d}m$ is the differential cross section (or more generally the underlying distribution for the training events), and a constant $C$ is required by dimensional analysis and should be common to signal and background samples.  In practice, we execute \eref{eqn:weight} by uniformly binning the input events and inverting the histogram, which introduces some finite bin effects as will be apparent below.   Note for a different purpose, the experimental collaborations frequently weight events  to match the transverse momentum spectrum of different samples (\emph{e.g}~\cite{Aad:2013wqa,Chatrchyan:2014nva, Aad:2015rpa}). 

Next, we train a new network on the planed input data.  The performance drop yields a measurement of the discriminating information contained in the variable $m$.  This procedure can be iterated, by choosing the next variable to plane with, until the network is unable to discriminate between the fully planed signal and background.  This end point demonstrates that all of the information available to distinguish signal from background is encoded in the planing variables, thereby providing a procedure to concretely frame the question posed by the title of this paper.

Planing is one of many different approaches to understanding a network's discrimination power as mentioned in the introduction and reviewed in~\cite{Larkoski:2017jix}.  In what follows, as we study planing we will also utilize a technique (see~\cite{Baldi:2014kfa,Baldi:2014pta,Baldi:2016fql,Guest:2016iqz,Datta:2017rhs,Aguilar-Saavedra:2017rzt}) which we refer to as \emph{saturation}, that compares a network trained on only low-level inputs with networks trained after adding higher-level variables.    Saturation provides a tool to  ensure that our networks are sufficiently deep, by checking that the new network's performance does not improve by much.\footnote{Saturation can also be used to determine what high-level variables provide information (\emph{e.g.} \cite{Datta:2017rhs,Aguilar-Saavedra:2017rzt}). Planing tests can be easier to interpret, due to its larger dynamic range in performance metrics.}   

Saturation additionally suggests another method to uncover what information a machine is utilizing.  One could consider training networks using only the high-level variable(s) of interest as inputs, where in contrast to the saturation technique, no low level information is being provided to the network. The diagnostic test would be to compute if the resulting network can achieve performance similar to that of a deep network that had been trained on only the low level inputs.  If the metrics were comparable, it would suggest that a machine can use the high-level variables alone to classify the data. However, the planing method has two advantages.  First, the number of input parameters would typically change when going from only low level to only high-level variables; unlike planing this requires altering the network architecture.  This in turn can impact the optimization of hyper-parameters, thereby complicating the comparison.  Furthermore, this method suffers the same issue as saturation in that as the limit towards ideal performance is achieved, one is forced to take seriously small variations in the metrics.  If there are not enough training trials to adequately determine the errors, these small variations could be incorrectly interpreted as consistent with zero.  This can again be contrasted with planing in that our approach yields a qualitative drop in performance and is more straightforward to interpret.

For all results presented below, we will distinguish the performance of a \emph{linear} versus \emph{deep} network.  This provides a diagnostic tool as to what extent the remaining information is a (non-)linear function of the inputs. The machines used here are neural networks implemented within the \texttt{Keras}~\cite{chollet2015keras} package with the \texttt{TensorFlow}~\cite{tensorflow2015-whitepaper} backend.  We choose either zero or three hidden layers to define the linear and deep networks respectively; each hidden layer has 50 nodes.  Note that a network with no hidden layer is equivalent to standard logistic regression. The inputs to each node are passed through the ReLu activation function, except that the Sigmoid is applied to the output layer.  Training is done using the Adam optimizer \cite{DBLP:journals/corr/KingmaB14}.  For each classification, $10\%$ of the events are used as a test set and $4.5\%$ are used for validation. Our metrics are computed on the test set using \texttt{scikit-learn}~\cite{scikit-learn}. We provide the standard metric for performance: the area under the receiver operating characteristic curve (AUC).  We compute standard deviation of the AUC by using the output of ten networks trained with randomly chosen initial conditions.  This is provided in the tables and gives a sense for the stability of the minimization.  

We will first demonstrate how to plane in a concrete toy example.  Assume the input data is given by three coordinates $\vec{x} = (x,y,z)$, and the signal is drawn from the distribution
\begin{align}
 f\big(\vec{x}\big) = 
\bigg[ \Theta\Big(r_0-r\,\Big) + C_r\bigg] \cdot \bigg[z \cdot B_z  +C_z\bigg]
\label{eq:toySignal}
\end{align}
where  $\Theta(x)$ is the step function, $r = \sqrt{x^2+y^2}$, the $C_i$ are constants, $r_0$ is the radius of a circular feature in the $x$-$y$ plane, and $B_z$ is the slope of the $z$-component of the signal.  The background distribution is uniform in $x$, $y$, and $z$.  This toy model contains both linear ($z$) and nonlinear ($x$-$y$) differences between signal and background, and it is also factorized such that there are no correlations between $r$ and $z$.

\begin{table}[htb]
\begin{center}
\renewcommand{\arraystretch}{1.5}
\setlength{\arrayrulewidth}{.4mm}
\centering
\setlength{\tabcolsep}{0.4em}
\begin{tabular}{ccc| c | c  }
$(x, y,z)$ & $r$ & \sc{Planed} & \sc{Linear AUC}& \sc{Deep AUC} \\
\hline
\ding{51} & \ding{56} & \ding{56} 	& $0.61275(01)$ 	& $0.81243(45)$	\\ 
\ding{51} & \ding{51} & \ding{56} 	&  $0.79672(01)$ 	& $0.81388(23)$ 	\\ 
\ding{51} & \ding{56} & $\bm{r}$	&  0.61030(01) 		& 0.61026(02) 		\\ 
\ding{51} & \ding{56} & $\bm{(r, z)}$ 	&  \!\!0.5081(16) 	& 0.49998(03)		\\ 
\end{tabular}
\end{center}
\vspace{-12pt}
\caption{The AUC output for a variety of input configurations applied to toy signal data pulled from \eref{eq:toySignal} and a flat background.  The variable $r$ is the  cylindrical radius.}
\label{tab:toyResults}
\end{table}%

The results of the study are presented in Table~\ref{tab:toyResults}.  First note that when training the networks on only the low level inputs, the deep network is more powerful.  This points to the presence of a nonlinearity, a consequence of the cylindrical shape of the underlying distribution.  Next, in the spirit of the saturation approach, we add the 2-$D$ radius $r$ to the list of inputs and train another network.  We see that the linear and deep networks perform nearly identically to the deep network trained only on the low-level inputs, which implies the remaining discriminating power is a linear function of the inputs, as it had to be from \eref{eq:toySignal}.  The third row shows the results when training on data whose $r$-dependence has been planed away.  All that remains is the $z$-dependence, which is linear as demonstrated by comparing the linear and deep outputs (see also Fig.~\ref{fig:LR_Correlation} left).  Finally, we plane in $r$ and $z$ simultaneously.  The bottom row of the table shows the AUC approaching $1/2$, signaling that all discriminating power is captured by $r$ and $z$.

\section{APPLICATION TO PARTICLE PHYSICS}
\label{sec:PhysicsExample}
\begin{figure*}[t]
\includegraphics[width=  0.85 \linewidth]{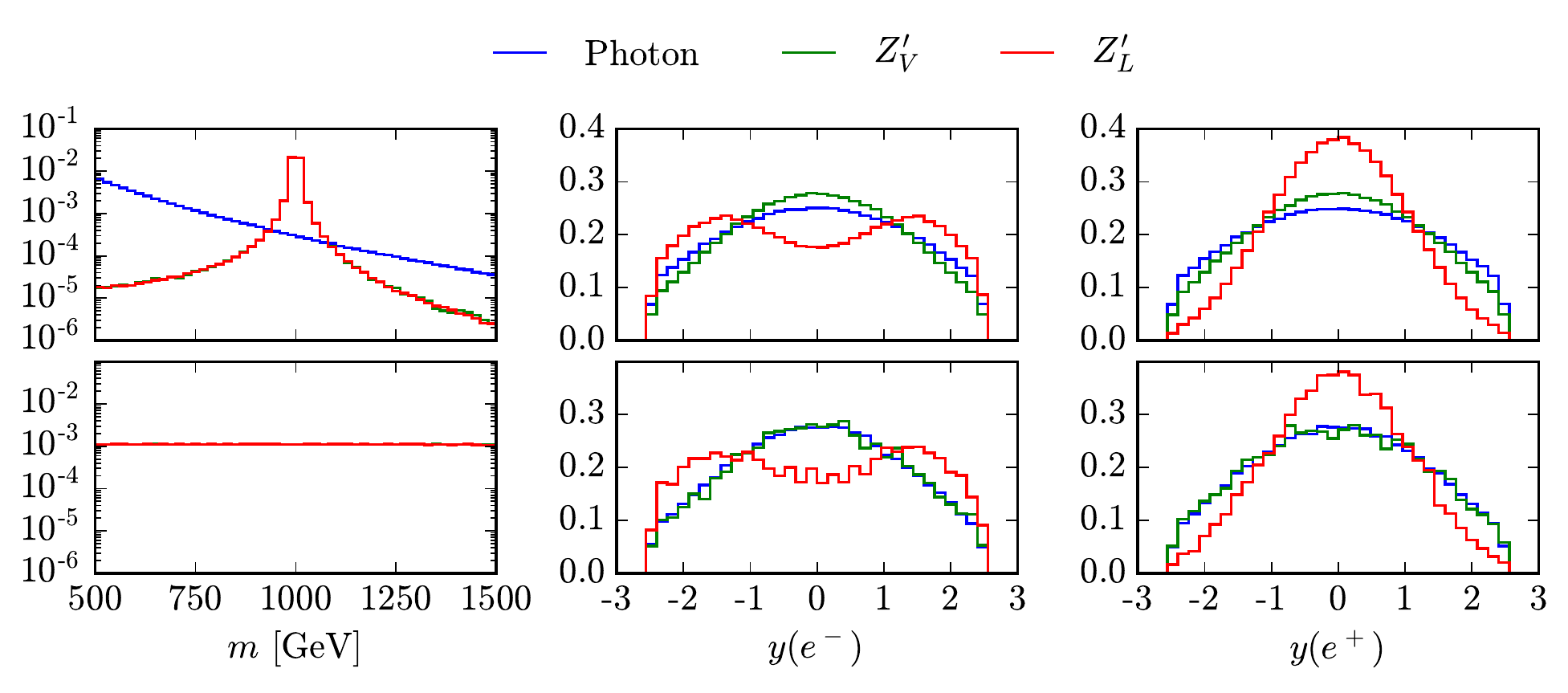}
\vspace{-10pt}
\caption{Histograms of the constructed variables normalized to unity. The top [bottom] panels are before [after] planing the input events using the invariant mass $m$. The rapidity of the electron (positron) is specified by $y(e^-)$ ($y(e^+)$).}
\label{fig:Distributions}
\end{figure*}

This section provides a planing application to a physical scenario.  We extend the Standard Model with a single particle, a massive vector boson $Z'$ that decays to an electron $(e^-)$ positron $(e^+)$ pair.  This example was chosen because the best discriminator against the smoothly falling photon background is the invariant mass $m^2 = (p_{e^+} + p_{e^-})^2$, a nonlinear combination of the input four-vectors $p$.  Furthermore, depending on how we choose the helicity structure of the coupling between the $Z'$ and the Standard Model particles, additional discriminating power beyond invariant mass may be present.  

We use a phenomenological parametrization:
\begin{align}
\mathcal{L} \supset Z'_\mu \sum_f Q_f \Big(g_{Z',L} \overline{f} \gamma^\mu P_L f + g_{Z',R}  \overline{f} \gamma^\mu P_R f \Big)\,,
\end{align}
where $f$ are the Standard Model fermions, $Q_f$ is the electric charge, $P_{L(R)}$ are the left (right) projection operators, $g_{Z',L(R)}$ is the strength of the coupling between the left (right) handed fermions and the $Z'$.  We take $M_{Z'} = 1\text{ TeV}$ and the width $\Gamma_{Z'} = 10 \text{ GeV}$.  This model is excluded by LHC data over a wide parameter space; we present it here solely as an instructive tool.

We will focus our attention on two cases:  $Z'_V$ with vector coupling where $g_{Z'L} = g_{Z'R}$ (the same as the helicity structure of the photon), and $Z'_L$ with left couplings active and $g_{Z'R}=0$.  The models are implemented using \texttt{FeynRules} \cite{Alloul:2013bka}.  
The Monte Carlo event generator \texttt{MadGraph} \cite{Alwall:2014hca} is used to simulate $10^6$ proton-proton collisions  with an invariant mass between 500 and 1500 GeV for $\gamma^{*}$, $Z'_V$, and $Z'_L$ intermediate states.  Using information contained in $p\,p \rightarrow e^+ e^-$ events, the goal is to distinguish the $Z^{\prime}$ signal models from the photon background. 

We take the low-level training inputs to be the four-vectors $(E, \vec{p}\,)$ of the $e^{\pm}$.  We know that the best discriminator between signal and background is the invariant mass.  This is the only distinguishing feature between the $Z'_V$ and the photon.  However, due to the nontrivial helicity structure of the $Z'_L$ model, there are additional features in the high-level variable rapidity, $y\equiv \frac{1}{2} \log[(E+p_z)/(E-p_z)]$, that distinguish it from the photon.
The distributions of the high-level variables are shown in the upper panels of Fig.~\ref{fig:Distributions}.  
 
The results of classifying the $Z^{\prime}_V$ against the photon are shown in Table~\ref{tab:PhotonVsVector}.  We train the linear and deep networks on the low-level variables, and again on the low-level variables plus invariant mass.  The deep network performance is very similar with or without the invariant mass; following the logic of the saturation approach, this shows that the low-level deep network is a nearly ideal discriminator. For comparison, the low-level linear network performance is far below that of the deep network.  We infer that nonlinear combinations of the input variables are needed to optimally classify the data.  When invariant mass is added to the linear network, the resulting performance significantly improves, but it still does not match the power of the deep networks.
One is tempted to (falsely) conjecture that there is extra discriminating power to uncover, and the top row of Fig.~\ref{fig:Distributions} seems to add support.   It is also possible that the linear network aided by $m$ does not perform as well as the deep network, even though it contains all of the relevant information, because it can only make a one-sided cut.

\begin{table}[h]
\begin{center}
\renewcommand{\arraystretch}{1.5}
\setlength{\arrayrulewidth}{.4mm}
\centering
\setlength{\tabcolsep}{0.4em}
\begin{tabular}{ccc| c | c  }
$(E, \vec{p}\,)$ & $m$ & \sc{Planed} & \sc{Linear AUC}& \sc{Deep AUC} \\
\hline
\ding{51} & \ding{56} & \ding{56} 	&	$0.746221(01)$ 	& $0.988510(98)$			\\ 
\ding{51} & \ding{51} & \ding{56} 	&   $0.938967(01)$ 		& $0.989007(03)$ 			\\ 
\ding{51} & \ding{56} & $\bm{m}$	& \!\!\!0.50550(29) 		& \hspace{-14pt} 0.4942(48) 	\\ 
\end{tabular}
\end{center}
\vspace{-12pt}
\caption{The AUC output for a variety of input configurations applied to the $Z^{\prime}_V$ model and the photon background. } 
\label{tab:PhotonVsVector}
\end{table}%

However, due to the vector nature of the photon couplings (and the masslessness of the final state particles), we know that the only difference between signal and background should be captured by the invariant mass of the electron positron pair.   To determine the correct interpretation, we plane signal and background in invariant mass as shown in the lower row of Fig.~\ref{fig:Distributions}. As expected, the photon and the vector $Z^{\prime}$ have nearly identical distributions up to the noise induced by the histograming procedure for computing the weights.

In order to quantify if there is information hidden in any of the other distributions, linear and deep networks are trained on the planed inputs. The results are shown in the lower section of Tab.~\ref{tab:PhotonVsVector} as measured on the planed test set. Both networks have an AUC approaching 0.5, so no noticeable discriminating power remains.  Since the planing process removed the  invariant mass information, the networks cannot tell the difference between the massless and massive vector boson propagators, showing that mass is in fact the only discriminator.  

\begin{table}[htb]
\begin{center}
\renewcommand{\arraystretch}{1.5}
\setlength{\arrayrulewidth}{.4mm}
\centering
\setlength{\tabcolsep}{0.4em}
\begin{tabular}{ccc| c | c  }
$(E, \vec{p}\,)$ & $m$ & \sc{Planed} & \sc{Linear AUC}& \sc{Deep AUC} \\
\hline
\ding{51} & \ding{56} & \ding{56} 				&$0.763280(05)$ 	& $0.989353(59)$			\\ 
\ding{51} & \ding{51} & \ding{56} 				&   $ 0.942004(02)$ 	& $ 0.989826(10)$ 			\\ 
\ding{51} & \ding{56} & $\bm{m}$				& 0.626648(28) 	&\hspace{-11pt}0.6258(24)	\\ 
\ding{51} & \ding{56} & $\bm{\,\,\,(m, \Delta|y|)\,\,\,}$ 	&  \!\!\!0.52421(15) 	& \hspace{-14pt} 0.5320(25) 	\\ 
\end{tabular}
\end{center}
\vspace{-12pt}
\caption{The AUC output for a variety of input configurations applied to the $Z^{\prime}_L$ model and the photon background.  The variable $\Delta |y| \equiv |y(e^-)|-|y(e^+)|$.}
\label{tab:PhotonVsPL}
\end{table}%

\begin{figure}[t]
\begin{center}
\hspace{-15pt} \includegraphics[width = 1.05 \linewidth]{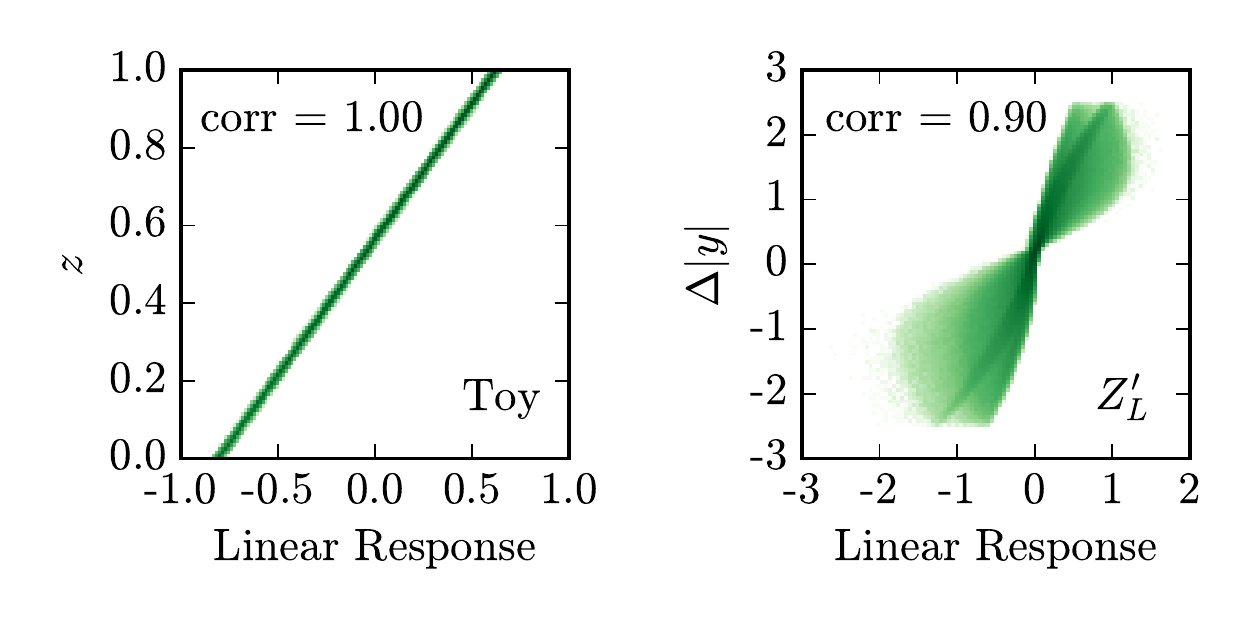}
\vspace{-25pt}
\caption{[Left] Density of events for the planed linear network output versus $z$ for the toy model presented in Sec.~\ref{sec:Proceedure}.  [Right] Density of events for the planed linear network output and $\Delta|y|$ for the $Z'_L$ model.  Both signal and background events are being plotted.  The correlation measure is provided in the top of each panel.  Perfect correlation would imply that the variable and linear network represent the same information.}
\vspace{-20pt}
\label{fig:LR_Correlation}
\end{center}
\end{figure}

Next, we explore the $Z'_L$ signal model where we expect additional discriminants to be present.  Networks are trained to distinguish the $Z'_L$ from the photon, with results shown in Table~\ref{tab:PhotonVsPL}.  Initially, we see a pattern similar to that as in the previous examples. Note that now the AUCs are slightly closer to unity as compared to the $Z^{\prime}_V$ model, again indicating the presence of information beyond the invariant mass.  An inspection of the distributions that have been planed using $m$, which are plotted in the lower panels of Fig.~\ref{fig:Distributions}, reveals the source of this additional discriminating power. The $Z^{\prime}_L$ clearly manifests differences in the rapidities for the electron and positron, where the magnitude of the electron rapidity is usually larger than the magnitude of the positron rapidity for the $Z^{\prime}_L$.  This results from the choice of chiral couplings and the shape of the parton distribution functions.  This suggests that a variable $\Delta |y| \equiv |y(e^-)| - |y(e^+)|$ should be a useful discriminator (the more traditional approach is to utilize asymmetry observables, \emph{e.g.} the reviews~\cite{Leike:1998wr, Rizzo:2006nw}). This can be further quantified by computing the correlation between the linear network response (before the Sigmoid activation) and $\Delta|y|$, as shown in the right panel of Fig.~\ref{fig:LR_Correlation}.  A correlation of $0.90$ is observed, implying that much of the remaining information is contained in $\Delta |y|$.  As a comparison, we also show the equivalent result derived for the toy model of Sec.~\ref{sec:Proceedure} in the left panel of Fig.~\ref{fig:LR_Correlation}.  Since the signal was linear in $z$ by construction, a perfect correlation is expected and demonstrated. Performing this test on any new variables is a powerful and quick method to assess their performance and test their linearity.

Next, we plane the inputs using the full $m$-$\Delta |y|$ dependence, and train new networks.  The results are provided in the last row of Table~\ref{tab:PhotonVsPL}.  We see that an AUC approaching $1/2$ is achieved for both the linear and deep networks.  The remaining bits of discriminating power could be resolved by planing in 3D: $(m,y(e^+), y(e^-))$.  This would determine to what extent it is due to physics as opposed to noise from the histograming procedure.  

\section{OUTLOOK}
\label{sec:Discussion}

We explored data planing, a probe of machine learning algorithms designed to remove features in a given variable, see also~\cite{deOliveira:2015xxd}.  By iteratively planing training data, it is possible to remove the machine's ability to classify.  As a by-product, the planed variables determine combinations of input variables that explain the machine's discriminating power.  This procedure can be explored systematically, but is most efficient in tandem with physics intuition.

In the future, it would be interesting to examine this procedure with more realistic training data that include initial/final state radiation and detector effects.  The application to more complicated signals should also be tested. With exotic signals, planing may need to be done in many dimensions;  perhaps a kernel smoothing procedure should be applied, or perhaps a network can be trained to compute the weights directly, which can then be utilized when training the planed network. Choosing which variables to plane in will be increasingly challenging in higher dimensional phase space, as highlighted in the example of jet images~\cite{deOliveira:2015xxd}.  Careful treatment of correlations will also be relevant; see~ \cite{Dolen:2016kst, Shimmin:2017mfk} for related ideas. 

One interesting extension would be to systematically test a large set of Lorentz invariants in order to find the combination that yields the largest performance drop. This could reveal new variables for traditional searches.  Finally, what information is contained in jets could be explored with planing to complement the existing saturation analyses \cite{Datta:2017rhs, Aguilar-Saavedra:2017rzt}.  We intend to investigate many of these applications in future studies.

\section*{Acknowledgements}
We would like to thank Marat Freytsis and Benjamin Nachman for useful comments on the draft.  T.C. is especially grateful to Ronnie Cohen for teaching him to use a plane in the real world.
This work is supported by the U.S. Department of Energy under Awardds No. DE-SC0011640 (to S.C. and B.O.) and DE-SC0018191 (to T.C.).

\begin{spacing}{1.1}

\bibliography{ML}

\end{spacing}

\end{document}